\documentclass[prd,twocolumn,aps,amsmath,nofootinbib,superscriptaddress]{revtex4}
\usepackage{epsfig}

  \newcount\hour \newcount\minute
  \hour=\time \divide \hour by 60
  \minute=\time
  \newcommand{\mydate}{\ \today \ - \number\hour :\ifnum \minute<10 0\fi 
\number\minute}

\newcommand{\nn}{\nonumber}

\newcommand{\OMIT}[1]{}

\begin{document}


\preprint{ \hbox{MIT-CTP 3574} \hbox{CALT-68-2534} \hbox{CMU-HEP-0411} 
 \hbox{hep-ph/04xxxxx}  }

\title{\boldmath
A New Method for Determining $\gamma$ from $B\to \pi\pi$ Decays 
}

\author{Christian W.~Bauer}
\affiliation{California Institute of Technology, Pasadena, CA 91125}
\author{Ira Z.~Rothstein}
\affiliation{Department of Physics, Carnegie Mellon University,
        Pittsburgh, PA 15213}
\author{Iain W.~Stewart\vspace{0.4cm}}
\affiliation{Center for Theoretical Physics, Massachusetts Institute of
  Technology, Cambridge, MA 02139}

\begin{abstract}
  \vspace{0.2cm}
  
  Factorization based on the soft-collinear effective theory (SCET) can be used
  to reduce the number of hadronic parameters in an isospin analysis of $B\to
  \pi\pi$ decays by one. This gives a theoretically precise method for
  determining the CP violating phase $\gamma$ by fitting to the $B\to \pi\pi$
  data without $C_{\pi^0\pi^0}$. SCET predicts that $\gamma$ lies close to the
  isospin bounds. With the current world averages we find $\gamma=75^\circ\pm
  2^\circ {}^{+9^\circ}_{-13^\circ}$, where the uncertainties are theoretical,
  then experimental.  

\end{abstract}

\maketitle

Measurements of CP violation are an important tool to look for physics beyond
the standard model (SM)~\cite{Zoltan}.  Standard model measurements of CP
violation in $B$-decays are usually expressed in terms of the angles $\alpha$,
$\beta$, $\gamma$. To test the SM picture of CP violation one looks for
inconsistencies by making measurements in as many decay channels as possible.

Important observables for measuring $\gamma$ (or $\alpha$) are the CP
asymmetries and branching fractions in $B \to \pi \pi$ decays.  Unfortunately,
hadronic uncertainties and ``penguin-pollution'' make the data difficult to
interpret.  Gronau and London (GL)~\cite{GL} have shown that using isospin,
${\rm Br}(\bar B \to \pi^+ \pi^-)$, ${\rm Br}(B^+ \to \pi^+ \pi^0)$, ${\rm
  Br}(\bar B \to \pi^0 \pi^0)$, and the CP asymmetries $C_{\pi^+\pi-}$,
$S_{\pi^+\pi-}$, $C_{\pi^0\pi^0}$, one can eliminate the hadronic uncertainty
and determine $\gamma$. Thus data is used to determine the 5 hadronic isospin
parameters. This year Babar and Belle~\cite{Cpi0pi0} reported a first
observation of $C_{\pi^0\pi^0}$. Unfortunately, the uncertainties in
$C_{\pi^0\pi^0}$ and ${\rm Br}(B\to\pi^0\pi^0)$ are still too large to give
strong constraints, leaving a four-fold discrete ambiguity and a $\pm 29^\circ$
window of uncertainty in $\gamma$ (at $1$-$\sigma$) near the SM preferred value.

In this letter we observe that the soft-collinear effective theory
(SCET)~\cite{SCET} predicts that one hadronic parameter vanishes at leading
order in a power expansion in $\Lambda_{\rm QCD}/E_\pi$, and that this provides
a robust new method for determining $\gamma$ using the experimental value of
$\beta$.  The parameter is $\epsilon = {\rm Im} (C/T)$, where $T$ and $C$ are
defined below and are predominantly "tree" and "color suppressed tree"
amplitudes.  From~\cite{bprs} we know that $\epsilon$ vanishes to all orders in
$\alpha_s(\sqrt{E_\pi\Lambda_{\rm QCD}})$ since the ``jet-function'' does not
involve a strong phase, and so $\epsilon$ receives corrections suppressed by
$\Lambda_{\rm QCD}/E_\pi$ or $\alpha_s(m_b)$.  Our method {\em does not rely} on
a power expansion for any of the other isospin parameters.  Thus, issues like
the size of charm penguins and whether ``hard-scattering'' or ``soft''
contributions dominate the $B\to \pi$ form
factors~\cite{cpens,BBNS,Keum,chay,bprs,Feldmann,Stan} are irrelevant here.  Our
analysis remains robust if so-called ``chirally enhanced'' power
corrections~\cite{BBNS} are included.  It differs from the QCDF~\cite{BBNS} and
pQCD~\cite{Keum} analyses; for example we work to all orders in $\Lambda_{\rm
  QCD}/m_b$ for most quantities and do not use QCD sum rule inputs.

The world averages for the CP averaged branching ratios ($\overline{\rm Br}$)
and the CP asymmetries are~\cite{Cpi0pi0,HFAG}\vspace{-0.2cm}
\begin{eqnarray}\label{data}
\mbox{
\begin{tabular}{l|ccc}
 & $\overline{\rm Br}\times 10^6$ & $C_{\pi\pi}$ & $S_{\pi\pi}$ \\\hline
$\pi^+ \pi^- $ &\ $4.6\pm 0.4$ \  & \ $-0.37\pm  0.11$ \   
 &\  $-0.61\pm 0.13$ \ \\
$\pi^0 \pi^0$ & $1.51\pm 0.28$ & $-0.28\pm  0.39$ & \\
$\pi^+ \pi^0$ & $5.61\pm 0.63$ & & 
\end{tabular}
}
\end{eqnarray}

\vspace{-0.2cm}
\noindent
For later convenience we define the ratios
\begin{eqnarray} \label{pipidata3}
  \overline R_c \!\!&=&\!\!
   \frac{\overline{\rm Br}(B^0\to \pi^+\pi^-)\tau_{B^-} }
   {2\overline{\rm Br}(B^-\to \pi^0\pi^-)\tau_{B^0}} = 0.446 \pm 0.064 \,,\nn\\
  \overline R_n \!\!&=&\!\! 
   \frac{\overline{\rm Br}(B^0\to \pi^0\pi^0)\tau_{B^-} }
   {\overline{\rm Br}(B^-\to \pi^0\pi^-)\tau_{B^0}} = 0.293 \pm 0.064\,,
\end{eqnarray}
and quote the product $\overline R_n\, C_{\pi^0\pi^0} = 0.082\pm 0.116$.

To obtain general expressions for these observables, we use isospin
and unitarity of the CKM matrix to write 
\begin{eqnarray}\label{b2pipi}
A(\bar B^0\to \pi^+\pi^-) &=& 
  e^{-i\gamma}\, |\lambda_u|\,   T - |\lambda_c|\, P\nn\\
 && 
  + \big(e^{-i\gamma}\, |\lambda_u|\!-\! |\lambda_c|\big) P_{\rm ew}^1 \,, \nn\\
%
A(\bar B^0\to \pi^0\pi^0) \!\!&=&\!\! 
  e^{-i\gamma}\, |\lambda_u| \, C + |\lambda_c| P \nn\\
 &&
 + \big(e^{-i\gamma}\, |\lambda_u|\!-\! |\lambda_c|\big) 
   (P_{\rm ew}^2 \!-\! P_{\rm ew}^1)\,, \nn\\
%
\sqrt2 A(B^-\to \pi^0\pi^-) \!\!&=&\!\! 
  e^{-i\gamma}\, |\lambda_u|\, (T +  C) \nn\\
 && 
   +\big(e^{-i\gamma}\, |\lambda_u|\!-\! |\lambda_c|\big) 
   P_{\rm ew}^2 \,.
\end{eqnarray}
Here $\lambda_u=V_{ub}^{\phantom{*}} V_{ud}^*$, $\lambda_c=V_{cb}^{\phantom{*}}
V_{cd}^*$.  The CP conjugate amplitudes are obtained from (\ref{b2pipi}) with
$\gamma\to -\gamma$.  $T$, $C$, $P$ and the electroweak penguin amplitudes
$P_{\rm ew}^{1,2}$ are complex.

The amplitude $P_2^{\rm ew}$ is related to $T$ and $C$ by isospin~\cite{su3}.
An additional relation for $P_1^{\rm ew}$ can be obtained using SCET at lowest
order in $\Lambda/E_\pi$ and $\alpha_s(m_b)$~\cite{bprs}.  For the
dominant coefficients $C_{9,10}$ we find
\begin{eqnarray}\label{eis}
  P_{\rm ew}^1 &=& e_1\: T + e_2 \: C\,,\ \ 
  P_{\rm ew}^2 =e_3\: T + e_4 \: C\,, \\[4pt]
  e_1 &=& \frac{ C_{10} (C_1\!-\! C_3) \!+\! C_9 (C_4\!-\! C_2)}
    {(C_1\!+\! C_2)(C_1\!-\!C_2 \!-\! C_3 \!+\! C_4)} 
    = -9.5\!\times\! 10^{-5}  \,,\nn\\
  e_2 &=& \frac{ C_{9} (C_1\!+\! C_4) \!-\! C_{10} (C_2\!+\! C_3)}
    {(C_1\!+\! C_2)(C_1\!-\!C_2 \!-\! C_3 \!+\! C_4)}  
    = -9.0\!\times\! 10^{-3}\,,\nn\\
  e_3 &=& e_4 = \frac{3}{2}\: (C_{9}\!+\! C_{10})(C_1\!+\! C_2)^{-1}
       = -1.4\times 10^{-2}\,,  \nn
\end{eqnarray}
with $C_i$ from the electroweak Hamiltonian.  Since $e_3 |T|/|P| = e_3
(p_s^2+p_c^2)^{1/2} |\lambda_c|/|\lambda_u| \sim 0.06$ for typical values of the
parameters $p_s$ and $p_c$ (from below), we estimate that the electroweak
penguins give at most a $\sim 6\%$ correction to any amplitude.  It would be
easy to include $P_{\rm ew}^{1,2}$, but for simplicity we neglect them in what
follows. SCET allows contributions from $C_7$ and $C_8$ to be included in
(\ref{eis}), giving $e_3=-1.5\times 10^{-2}$, $e_4=-1.3\times 10^{-2}$.

\begin{figure}[t!]
  \centerline{ 
  \raisebox{0.2cm}{\epsfysize=2.2truecm \hbox{\epsfbox{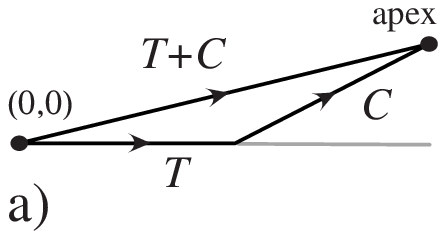}} }
   \hspace{-0.2cm} 
   \mbox{\epsfysize=2.4truecm \hbox{\epsfbox{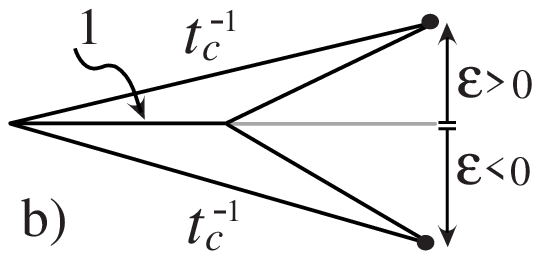}} } }
  \vskip-0.4cm {\caption[1]{a) Isospin triangle in $|\lambda_u|$ sector, and b) the
      rescaled triangle with solutions for positive and negative $\epsilon$ shown.}
\label{fig_isospin} }
\vskip -0.3cm
\end{figure}

Of the five remaining isospin parameters, one, $|\lambda_u(T+C)|$, is fixed by
$\overline {\rm Br}(B^-\to \pi^0\pi^-)$ and just sets the overall scale. We choose
the remaining four parameters as
\begin{eqnarray} \label{param2}
  p_c \!\! &\equiv& \!
     - \frac{|\lambda_c|}{|\lambda_u|}\: {\rm Re}\Big(\frac{P}{T}\Big)\,, 
  \quad
  p_s \!\equiv 
     - \frac{|\lambda_c| }{|\lambda_u|}\: {\rm Im}\Big(\frac{P}{T}\Big) \,,
   \nn \\[3pt]
  t_c \!\!&\equiv&\!  \frac{|T|}{|T+C|}  \,, 
  \qquad\qquad
  \epsilon \!\equiv {\rm Im}\Big(\frac{C}{T}\Big) \,. 
\end{eqnarray}
The parameters $p_c$ and $p_s$ determine the size of the ``penguin''
contribution $P$ relative to the ``tree'' $T$, and the parameters $t_c$ and
$\epsilon$ determine the shape of the isospin triangle as shown in
Fig.~\ref{fig_isospin}. Relations to parameters used
previously~\cite{pipianalysis,bprs} are $r_c^2=p_c^2\!+\!p_s^2$ and
$\tan\delta_c=p_s/p_c$.

In terms of the parameters in (\ref{param2}) the observables can be written as
(neglecting electroweak penguins)
\begin{eqnarray} \label{rslt1}
 S_{\pi^+\pi^-} \!\!&=&\!\!
    -\big[ \sin(2\beta\!+\!2\gamma)
     \!+\! 2 \sin(2\beta\!+\!\gamma) p_c\nn \\
  &&\!\!\!
  +\sin(2\beta)(p_c^2\!+\!p_s^2) \big] \big[ 1\!+\! 2 p_c \cos\gamma 
  \!+\! p_c^2 \!+\! p_s^2 \big]^{-1} 
      \!\! ,\nn\\[3pt]
 C_{\pi^+\pi^-} \!\!&=&\!\!
     \frac{2 p_s \sin\gamma}{1\!+\! 2 p_c \cos\gamma \!+\! p_c^2 \!+\! p_s^2} 
     \,,\nn\\[3pt]
&&\hspace{-1.45cm}
\overline R_c =
   {t_c^2} ( 1 + 2 p_c \cos\gamma + p_c^2 + p_s^2)
      \,,\nn\\[3pt]
&&\hspace{-1.5cm}
 \overline R_n =
   (1\!-\!t_c)^2 \!+\! t_c^2 (p_c^2\!+\!p_s^2)
    \!-\! {2 t_c (1\!-\!t_c) p_c } \cos\gamma \nn\\
 &&
 \hspace{-1cm}
  - \epsilon\, (2\, t_c^2\, p_s) \cos\gamma 
  + \Big[1\!\mp\!\sqrt{1\!-\!\epsilon^2 t_c^2}\Big] 2 t_c (1 \!+\! p_c \cos\gamma)
 \,, \nn \\
&&\hspace{-1.5cm}
\overline R_n C_{\pi^0\pi^0} =
   2 t_c \sin\gamma \Big[ t_c p_s \!\mp \!p_s \sqrt{1\!-\!\epsilon^2t_c^2}
   +\epsilon\, p_c t_c \Big].
\end{eqnarray}
The $\mp$ signs in the last two equations should be chosen to be the same, and
correspond to whether the apex of the triangle in Fig.~\ref{fig_isospin}a) is to
the right/left of the $(0,0)$ point.  Since both $|\lambda_c|$ and $|\lambda_u|$
are absorbed into the hadronic parameters $p_s$ and $p_c$, so there is no added
uncertainty from $|V_{ub}|$. For the CKM angle $\beta$ we use the latest
average~\cite{Zoltan,HFAG}, $\beta = 23.3^\circ\pm 1.5^\circ$.

\begin{figure}[t!]
  \centerline{ \mbox{\epsfxsize=7.7truecm \hbox{\epsfbox{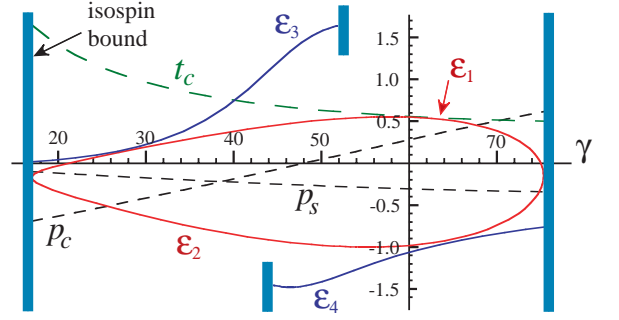}} } }
  \vskip-0.2cm {\caption[1] {Isospin analysis showing the hadronic parameters
      $\{p_c,p_s,t_c,\epsilon\}$ versus $\gamma$ using current central values of
      the $B\to \pi\pi$ data. Solutions for $\gamma$ occur at crossings of the
      $\epsilon_i$ curves. Experimental uncertainties are not shown, and are
      especially large for $\epsilon_{3,4}$. This plot shows only one of two
      allowed $(p_c,p_s)$ solutions and one of the two allowed
      $\gamma$-regions. }
\label{fig:fulliso} }
\vskip -0.3cm
\end{figure}

The full isospin analysis requires solving the five equations (\ref{rslt1}) to
obtain the parameters $p_c$, $p_s$, $t_c$ and $\epsilon$ defined in
(\ref{param2}) and the weak angle $\gamma$. From $S_{\pi^+ \pi^-}$ and $C_{\pi^+
  \pi^-}$ one obtains two solutions for the parameters $p_s$ and $p_c$ as
functions of the angle $\gamma$.  Using these, $\overline R_c$ determines
$t_c(\gamma)$.  Finally, $\overline R_n$ and $\overline R_n C_{\pi^0 \pi^0}$
each give two quadratic equations for $\epsilon$, which in general have four
intersections in the $\epsilon-\gamma$ plane.  We call $\epsilon_{1,2}$ the two
solutions from $\overline R_n$ and $\epsilon_{3,4}$ the two solutions from
$\overline R_n C_{\pi^0 \pi^0}$.  An example of this GL isospin analysis is
shown in Fig.~\ref{fig:fulliso}, where we use the current central values for the
data.  For illustration we picked the solution for $p_c$ and $p_s$ with
$|P/T|<1$, but have shown all four $\epsilon_i$'s.

An obvious feature in Fig.~\ref{fig:fulliso} are the isospin bounds on $\gamma$.
It is well known that there are bounds on $\gamma$ in the absence of a
measurement of $C_{\pi^0 \pi^0}$~\cite{isospinbound}.  To find these
analytically, one defines $\gamma=\pi-\beta-\alpha_{\rm eff}+\theta$
where
\begin{eqnarray} \label{GLSSbnd}
  \sin(2\alpha_{\rm eff}) \!\!&=&\!\! 
    S_{\pi\pi}(1\!-\!C_{\pi\pi}^2)^{-1/2}
    = -0.66 \pm 0.14\,, \\
  \cos(2\theta) \!\!&\ge&\!\! (\overline R \!-\!1)
   (1\!-\!C_{\pi\pi}^2)^{-1/2} = 0.53 \pm 0.19 \,,\nn
\end{eqnarray}
with $\overline R= (1\!+\! \overline R_c \!-\! \overline R_n)^2/(2\overline
R_c)$. The four solutions are
\begin{eqnarray} \label{bnds}
 -163.^\circ \!\!&\le&\! \gamma\:\, \le -105.^\circ \,,\quad
  -31.8^\circ \le \gamma\:\, \le\:\: 26.3^\circ \,, \nn\\
 17.1^\circ \!\!&\le&\! \gamma\:\, \le\:\: 75.2^\circ \,,\quad\ \ 
  148.^\circ \le \gamma\:\, \le\:\: 206.^\circ \,,
\end{eqnarray}
with uncertainty $\pm 8.2^\circ$ on each limit.  At each of these bounds the two
solutions $\epsilon_{1,2}$ become degenerate and beyond they are complex,
indicating that the isospin triangle does not close.

Solutions for $\gamma$ from the GL isospin analysis are given where the curves
$\epsilon_{3,4}$ intersect the curves $\epsilon_{1,2}$. There are up to four
solutions within each isospin bound, which are symmetric around $\gamma_{\rm
  eff}=\pi-\beta-\alpha_{\rm eff}$. We show in Fig.~\ref{fig:fulliso} the
results for $17.1^\circ\le \gamma\le 75.2^\circ$.  The current central values
$\epsilon_{3,4}$ do not intersect $\epsilon_{1,2}$, and in the absence of
experimental uncertainties there would be no solution for $\gamma$. The current
central values for the observables are such that the solutions for $\epsilon$
from $\overline R_n$ and $\overline R_n C_{\pi^0 \pi^0}$ are almost tangential.
Including experimental uncertainties a large range of $\gamma$ is allowed, with
the highest confidence at $\gamma = 27^\circ$ and $\gamma = 65^\circ$.  This
conclusion agrees with the CKMfitter group's analysis which incorporates
$C_{\pi^0\pi^0}$~\cite{CKMfitter}.

Using SCET at LO in $\alpha_s(m_b)$ and $\Lambda_{\rm QCD}/m_b$ we have
$\epsilon=0$~\cite{bprs}, which corresponds to flat isospin triangles in
Fig.~\ref{fig_isospin}. Equivalently
\begin{eqnarray} \label{epspc}
 \epsilon\sim {\cal O}\Big(\frac{\Lambda_{\rm QCD}}{m_b}\,, \alpha_s(m_b)\Big) \,.
\end{eqnarray}
Neglecting EW-penguins, $\epsilon$ is an RGE invariant quantity since
Eq.~(\ref{rslt1}) fixes it in terms of observables.  Eq.~(\ref{epspc}) makes an
extraction of $\gamma$ from $B\to \pi\pi$ possible without needing precision
data on $C_{\pi^0\pi^0}$.  In this method the central values for $\gamma$ are
determined by finding the places where the $\epsilon_1$ and/or $\epsilon_2$
curves cross the x-axis, meaning we solve $\epsilon_{1,2}(\gamma)=0$. The other
hadronic parameters, $p_c$, $p_s$, and $t_c$ are determined in the same way as
in the isospin analysis. This proposal for determining $\gamma$ using
Eq.~(\ref{epspc}) is the main result of this letter. 

Using the central values for all the data besides $C_{\pi^0\pi^0}$ and solving
$\epsilon_{1,2}(\gamma) = 0$ gives the solutions
\begin{eqnarray}\label{gammacentral}
 \gamma = -159^\circ\,,\ \ -105^\circ\,,\ \  21.5^\circ\,,\ \  74.9^\circ \,.
\end{eqnarray}
We have four solutions rather than the eight of the isospin analysis (which
occur within the first and third isospin bounds in (\ref{bnds})), because
factorization for the $B\to \pi\pi$ amplitudes resolves the discrete ambiguity
in $p_s$ and $p_c$ in favor of $|P/T|<1$ solutions (this follows from the
factorization for light-quark penguins, the size of Wilson coefficients, charm
velocity power counting, and factors of $\alpha_s(m_c)$~\cite{BBNS,bprs}). Next
we analyze the theoretical and experimental uncertainties in our method for
$\gamma$, and contrast these with the isospin analysis, focusing on the two
solutions which can occur in the $17.1^\circ \le \gamma \le 75.2^\circ$ region
preferred by global fits for the unitarity triangle~\cite{CKMfitter}.

To estimate the theoretical uncertainty we take
\begin{eqnarray}
-0.2 \le \epsilon \le 0.2\,,
\end{eqnarray}
which corresponds to roughly a 20\% effect from perturbative or power
corrections. We also consider a much more pessimistic scenario where this range
is doubled to $\epsilon=\pm 0.4$.  Note that $|\epsilon| < 0.2$ can accommodate
the so-called ``chirally enhanced'' power corrections, which have been argued to
dominate~\cite{BBNS}. Using the results from Ref.~\cite{BBNS}, including known
$\alpha_s(m_b)$ and power corrections, we randomly scan the two complex
parameters $X_A$ and $X_H$ in the BBNS range to find $|\epsilon|=|{\rm
  Im}\big({C}/{T}\big)|_{\rm QCDF} = -0.05\pm 0.04$.  This is below the
uncertainty assigned to our analysis.

\begin{figure}[t!]
  \centerline{\mbox{\epsfxsize=7.5truecm \hbox{\epsfbox{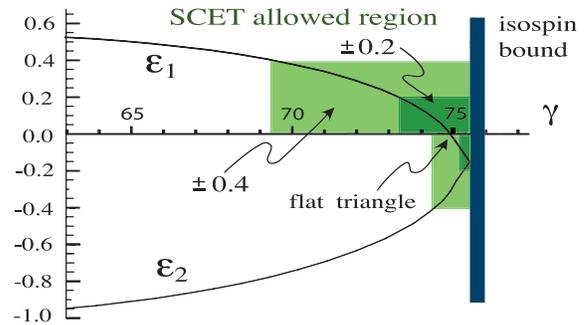}} } }
  \vskip-0.2cm {\caption[1] {Regions of $\gamma$ preferred by our analysis.
      The shaded areas show our best estimate of the theoretical uncertainty
      from power corrections, $-0.2 \le \epsilon \le 0.2$ as well as the
      pessimistic estimate $-0.4\le \epsilon \le 0.4$.  Experimental
      uncertainties are not shown.}
\label{fig:errors} }
\vskip -0.3cm
\end{figure}
In Fig.~\ref{fig:errors} we show $\epsilon_{1,2}$ for the region $65^\circ <
\gamma < 78^\circ$. Here the solution is $\gamma=74.9^\circ$, and the different
shading corresponds to the theory uncertainty with $|\epsilon| < 0.2(0.4)$.  The
solution for $\gamma$ is very close to the isospin bound, so the upward
uncertainty on $\gamma$ is very small. (The uncertainty in the isospin bound is
contained in the experimental uncertaintiy.) For the downward uncertainty we
consider the overlap with the shaded region. For $|\epsilon| < 0.2$ we find $
\Delta\gamma_{\,\rm theo} =^{+0.3^\circ}_{-1.5^\circ}$, while for $|\epsilon| <
0.4$ we find $\Delta\gamma_{\,\rm theo} = ^{+0.3^\circ}_{-5.2^\circ}$. On top of
that there are uncertainties from hadronic isospin violation, typically
$\lesssim 3\%$, which we take to be $\pm 2^\circ$ since no source of enhanced
isospion violation has been identified. (A slightly larger $\sim 5^\circ$
uncertainty was found in~\cite{Gardner}, but using a smaller penguin amplitude.
Larger isospin violation can be accounted for by scaling up the $\pm 2^\circ$
lower bound on our theory error.)  Thus, with perfect data at the current
central values we arrive at a theory uncertainty with $|\epsilon| < 0.2$ as
$\Delta\gamma_{\,\rm theo} = \pm 2^\circ$.
Repeating for the smaller solution at $\gamma=21.5^\circ$, we find a larger
theory uncertainty, $\Delta\gamma_{\,\rm theo} = {}_{-4.4^\circ}^{+8.7^\circ}$,
since the $\epsilon_{1,2}$ curves are flatter near this solution.

To determine the 1-$\sigma$ experimental errors, we use the program Minuit.
Taking $\epsilon=0$ and fitting to $\gamma$ and the four hadronic parameters we
find
\begin{eqnarray}
 \gamma = 21.5^\circ {}^{+9.4^\circ}_{-7.9^\circ}\,,\qquad
 \gamma = 74.9^\circ {}^{+8.1^\circ}_{-10.6^\circ} \,. 
\end{eqnarray}
These uncertainties are purely experimental and are propagated with the
assumption that the original input data are uncorrelated. If we instead set
$\epsilon=0.2$ then we find $ \gamma = 73.3^\circ {}^{+8.8^\circ}_{-13.3^\circ}$
and $\gamma = 30.7^\circ{}^{+11.1^\circ}_{-7.2^\circ} $, whereas fixing
$\epsilon=-0.2$ gives $ \gamma = 75.2^\circ {}^{+7.6^\circ}_{-9.5^\circ}$ and
$\gamma = 17.2^\circ {}^{+8.7^\circ}_{-6.9^\circ} $.  Combining these numbers we
obtain our final result for $\gamma$ including all sources of uncertainty
\begin{eqnarray} \label{sl1}
  \gamma = 74.9^\circ \pm 2^\circ {}^{+9.4^\circ}_{-13.3^\circ}    \,.
\end{eqnarray}
Here the first error is theoretical, and the last errors are experimental where
we picked the largest range obtained in varying $\epsilon= \pm0.2$. The theory
error increases to $\Delta \gamma={}^{+2^\circ}_{-5.2^\circ}$ for the more
pessimistic case $\epsilon=\pm 0.4$.  The analog of (\ref{sl1}) for the lower
solution is $ \gamma = 21.5^\circ {}^{+8.7^\circ}_{-4.4^\circ}
{}^{+11.1^\circ}_{-7.9^\circ}$.

The analysis presented here relies on the fact that a small value of
$|\epsilon|$ is allowed only for a narrow range of $\gamma$. While this is
certainly true given the current central values of the data, it is instructive
to investigate how the quality of the analysis is affected if the data central
values change. For example, it could be that the value of $\epsilon_1$ never
exceeds $0.2$, increasing the uncertainties from the small-$\epsilon$ analysis
significantly. A second extreme situation is that $\epsilon$ never reaches zero.
To study these questions, we generate random sets of data using Gaussian
distributions with the current central values and width of the 1-$\sigma$
uncertainties. We generate 10000 sets of ``data'', and after imposing $\sin(2
\alpha_{\rm eff}) < 1$ and $\cos (2 \theta)<1$ are left with 9688 sets.  Of
these 96\% have solutions for $\epsilon = 0$.  For $\epsilon_1$ we find 88\%
(70\%) of the sets have the maximum value above 0.2 (0.4).  It is only for these
data sets that our analysis works. For $\epsilon_2$ we find that $\sim 100\%$ of
the sets have their minimum below -0.4. Thus, the small $\epsilon$ analysis
works well in most cases.

\begin{figure}[t!]
  \centerline{ \mbox{\epsfxsize=9.3truecm
      \hbox{\epsfbox{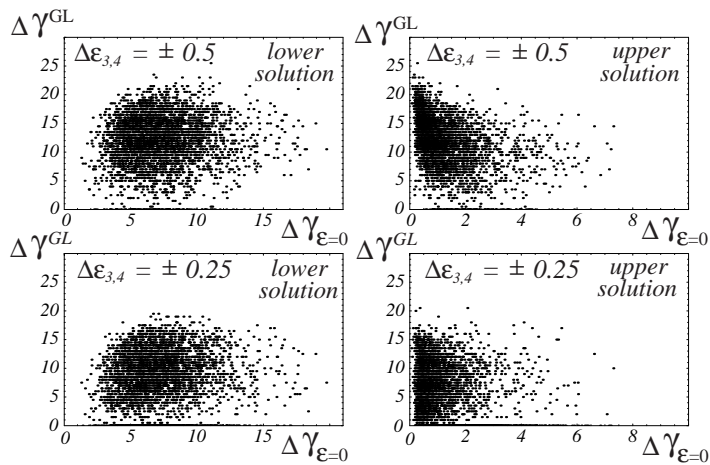}} } } 
 \vskip-0.3cm 
 {\caption[1] {Uncertainty in the isospin analysis from $C_{\pi^0\pi^0}$
     (y-axis) vs. theoretical uncertainty from our new method (x-axis). The
     upper (lower) two plots use a $\pm 0.5$ ($\pm 0.25$) uncertainty in
     $\epsilon_{3,4}$. The plots on the left (right) correspond to the solution
     near the lower (upper) isospin bound. }
\label{fig:compare} }
\vskip -0.4cm
\end{figure}
We can also study the uncertainty in our analysis, compared to the GL isospin
analysis. Rather than performing a full error analysis for the 9688 sets, we use
the following approximation. We assume that experimental uncertainty $\Delta
C_{\pi^0 \pi^0}$ dominates, and compare the resulting uncertainty in the GL
isospin analysis to the theoretical uncertainty in our analysis, for cases where
values of $\gamma$ exist with $\epsilon>0.2$ as discussed above.  The current
$\Delta C_{\pi^0 \pi^0}=\pm 0.39$ gives rise to a $\Delta\epsilon_{3,4}\sim \pm
0.5$.  In Fig.~\ref{fig:compare} we show the uncertainties in the GL analysis
compared with the theoretical uncertainties of the analysis presented here, for
both solutions of $\gamma$. The plots use 4000 points. If we take
$\Delta\epsilon_{3,4}\sim \pm 0.25$, the GL analysis still has uncertainties in
$\gamma$ that are considerably larger than the small $\epsilon$ analysis. We
also see that the $\pm 2^\circ$ error quoted in (\ref{sl1}) is typical.

We have presented a new method for obtaining $\gamma$ from $B \to \pi \pi$
decays without $C_{\pi^0 \pi^0}$. Our analysis uses SCET to eliminate one
hadronic parameter. The theory uncertainty for a solution $\gamma=74.9^\circ$
are small, $\pm 2$ or ${}^{+2^\circ}_{-5.2^\circ}$, depending on the estimate
for power corrections.  Analyzing possible future shifts in the data and
decreases in the $C_{\pi^0 \pi^0}$ uncertainty, we find that this method should
have smaller uncertainty than the isospin analysis for quite some time. The
analysis can be redone including the electroweak penguins. Obviously agreement
between Babar and Belle on $S_{\pi+\pi-}$, $C_{\pi^+\pi^-}$ is needed before
one will have complete trust in the $\gamma$ from $B\to\pi\pi$.

We would like to thank D.Pirjol for collaboration in early stages of this paper
and Z.Ligeti and J.Zupan for comments on the manuscript. This work was supported
in part by the DOE under DE-FG03-92ER40701, DOE-ER-40682-143, DEAC02-6CH03000,
the cooperative research agreement DF-FC02-94ER40818. I.S. is also supported by the
Office of Nuclear Science, a DOE OJI award, and a Sloan fellowship.


\end{document}